\documentclass[12pt]{iopart}
\usepackage{graphicx, epsfig}
\begin{document}

\title[Centrality and rapidity dependence of particle ratios in Au+Au and Cu+Cu ...]{Centrality and rapidity dependence of particle ratios in Au+Au and Cu+Cu 
collisions at $\sqrt{s_{NN}}$ = 62.4 GeV}

\author{I.C.Arsene\footnote{Also at Institute for Space Sciences, Bucharest, Romania} for BRAHMS collaboration}

\address{Department of Physics, University of Oslo, Norway}
\ead{i.c.arsene@fys.uio.no}
\begin{abstract}
We report on preliminary identified particle ratios from  Au+Au collisions 
at $\sqrt{s_{NN}} = 62.4$ GeV in different centrality classes, measured with the BRAHMS spectrometer. 
Results from Cu+Cu and p+p collisions at mid-rapidity at the same energy are also included.
The average transverse momenta of particle spectra, anti-particle to particle ratios and $K/\pi$ 
ratios dependence on centrality and rapidity are shown and discussed.
\end{abstract}



\section{Introduction}
Particle ratios are an important tool for studying the particle production
in nucleus-nucleus collisions. Their dependence on rapidity and centrality gives an insight into the 
reaction dynamics and production mechanisms. 
At RHIC energies, the anti-particle to particle ratios in Au+Au collisions at mid-rapidity 
indicate the formation of a highly matter/anti-matter equilibrated system \cite{brahmsRat}.
At forward rapidity, where the net-baryon content is significant \cite{brahms62Stopping}, these ratios fall
due to increasing contribution from production mechanisms other than the pair production.
The centrality dependence of the anti-particle to particle ratios shows that there is a 
smooth transition from peripheral nucleus-nucleus collisions towards the results from 
p+p collisions \cite{starReview}.

The ratio of strange to non-strange particles measured in all A+A collisions shows an enhancement compared
to the N+N reactions considered at the same energy. This enhancement depends on energy \cite{agsKpiEnergy,spsKpiEnergy}, 
system size \cite{agsKpiSystemSize,spsKpiSystemSize}, 
and collision geometry \cite{spsGeometry}. 
In this work we will concentrate on the dependence of the average transverse momenta, 
ratios of yields of anti-hadrons to hadrons ($\pi^{-}/\pi^{+}$, $K^{-}/K^{+}$,
and $\bar{p}/p$), and $K/\pi$ ratios on the system size and geometry at mid- and forward rapidity. 
We will use data from Au+Au, Cu+Cu and p+p collisions at $\sqrt{s_{NN}}=62.4$ GeV.

\section{Experimental setup and data analysis}
The BRAHMS \cite{brahmsNIM, brahmsWhite} experiment consists of two rotatable spectrometer arms for 
particle identification and momentum measurement. It also contains a set of global detectors 
for global event characterization. 
BRAHMS has a very wide phase space coverage with very good particle
identification using time-of-flight walls and a ring imaging Cherenkov detector. 
In the $\sqrt{s_{NN}} = 62.4$ GeV dataset BRAHMS identified charged 
particles up to rapidity 3.4 which is less than one rapidity unit away from the beam rapidity ($y_{beam}=4.2$). 
The $\pi/K$ separation can be made up to $p\approx2.5$ GeV/$c$ near mid-rapidity and 
$p\approx20$ GeV/$c$ at $y>2$. The protons are well identified up to $p\approx3.0$ GeV/$c$ near mid-rapidity 
and $p\approx30$ GeV/$c$ at $y>2$.
The centrality is calculated by measuring the charged particle multiplicity in the pseudorapidity
interval $|\eta| < 2.2$. For that we use the Multiplicity Arrays which consists of two sets of detectors, 
silicon and scintillator tiles, placed on two barrels around the beam pipe at the nominal interaction point.

The spectra were corrected for tracking and PID efficiencies, acceptance, in-flight
weak decays, multiple scattering and hadronic interactions. 
The invariant $p_{T}$ spectra were
extracted in different rapidity windows and fitted with several theoretically motivated functions
in order to extract the integrated yields. The charged pion spectra were fitted  
with a power law function of the form $A(1+p_T/p_0)^{-B}$
while the kaon and proton spectra were fitted with an $m_T$ exponential function
\cite{auau62Yields}.
The average transverse momentum was calculated using the fitted functions and the
resulting $p_T$ integrated yields were used to calculate the anti-particle to particle
ratios and the $K/\pi$ ratios.

Two transport models were employed for comparison of theoretical calculations with the data, namely
AMPT \cite{ampt} and UrQMD \cite{urqmd}. Both of the models include a partonic initial stage
followed by a Lund-type string fragmentation and hadronic rescatterings. 

\section{Results and discussions}

\begin{figure}[hpt]
\begin{center}
\includegraphics[height=3.5in]{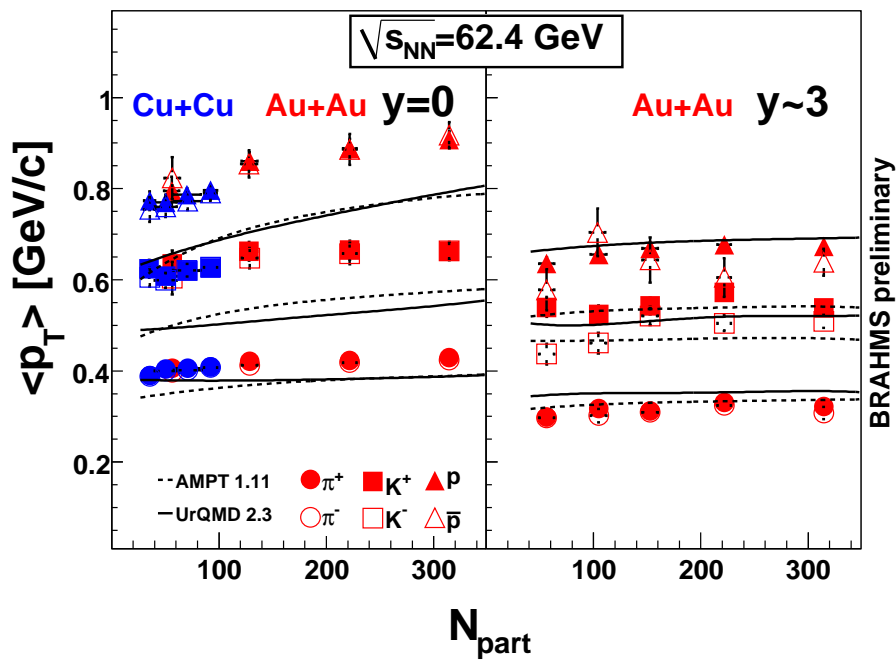}
\end{center}
\caption{Average $p_T$ for identified particles as a function of the number of participants. 
The data at mid-rapidity (left panel) is from Au+Au (red) and Cu+Cu (blue) collisions at $\sqrt{s_{NN}}=62.4$ GeV
and the data at forward rapidity (right panel) is from Au+Au collisions only. The curves are model calculations
with AMPTv1.11 \cite{ampt} (dashed line) and UrQMDv2.3 \cite{urqmd} (solid line). The curves
respect the same mass hierarchy as in the data.}
\label{avpt}
\end{figure}

In Fig.\ref{avpt} we show the centrality dependence of the average transverse momentum for the
identified particles at $y=0$ (left) and forward rapidity (right). 
The number of participants nucleons ($\mathrm{N_{part}}$) are estimated using Glauber HIJING calculation 
as described in \cite{brahmsCent}.
At mid-rapidity, in the
$\mathrm{N_{part}}$ range probed by using Au+Au and Cu+Cu collisions, 
the $\langle p_T\rangle$ increases with increasing the system size and centrality of the collision.
This together with the particle mass ordering suggests the presence of effects like 
the transverse radial flow due to the pressure developed during the collision. 
There are however other effects which can contribute also to the observed values of the transverse momenta
like jets, $k_{\perp}$ broadening or rescatterings in the hadronic phase.
At forward rapidity, $y\sim3$, much smaller values of the $\langle p_T \rangle$ were measured especially for
protons and kaons but the same mass ordering as in mid-rapidity is kept. Also, there is no or 
little dependence of $\langle p_T \rangle$ on centrality in contrast to mid-rapidity.
The curves in Fig.\ref{avpt} represent the theoretical calculations with AMPT (dashed lines)
and UrQMD (solid lines). The results of the calculations keep the same mass ordering as in the data
both at mid-rapidity (left panel) and forward rapidity (right panel). 
At mid-rapidity, both models give the right system size dependence but
underestimate the absolute values of the average transverse momentum. The calculations at
forward rapidity shows a better agreement with the data with the exception of the proton
average $p_T$ calculated with AMPT, which underpredicts the results.

\begin{figure}[hpt]
\begin{center}
\includegraphics[height=3.5in]{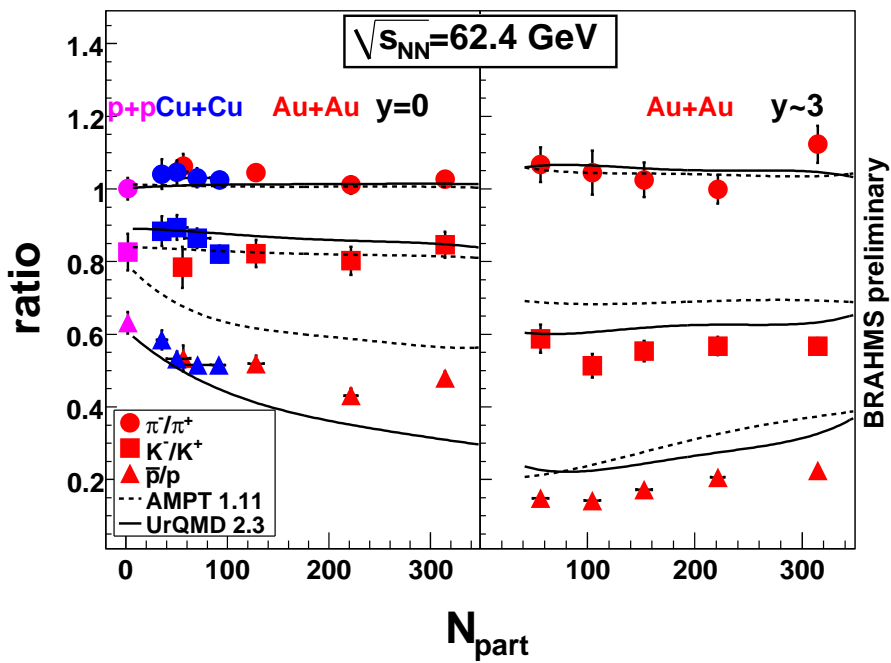} 
\end{center}
\caption{Anti-particle to particle ratios as a function of the number of participants at mid-rapidity (left panel)
and forward rapidity (right panel). The data at mid-rapidity includes BRAHMS results from Au+Au (red), Cu+Cu (blue) 
and p+p (magenta) collisions at $\sqrt{s_{NN}}=62.4$ GeV. At forward rapidity only Au+Au results are used.
The curves are model calculations using AMPT (dashed line) and UrQMD (solid line).}
\label{likeRat}
\end{figure}

The chemical conditions of the system can be studied with the help of anti-particle to particle ratios of the integrated $dN/dy$ yields.
The errors due to efficiencies are cancelled in these ratios while the systematic errors due to extrapolations
at low $p_T$ approximately cancel also. The mid-rapidity anti-particle to particle ratios are shown in
the left panel of Fig.\ref{likeRat}. 
The $\pi^{-}/\pi^{+}$ ratio is approximately 1 independent on centrality or reaction. 
Within the statistical errors, the measured $K^{-}/K^{+}$ ratio stays constant with centrality.
The $\bar{p}/p$ ratio shows a significant decrease from p+p results towards central Au+Au
which is consistent with larger baryon stopping in central nucleus-nucleus collisions.
The theoretical calculations made with the UrQMD and AMPT models agree with the data for $\pi^{-}/\pi^{+}$
and $K^{-}/K^{+}$ ratios. The calculations for $\bar{p}/p$ ratio seem to follow the trend of the data
but the absolute values are not well understood.

In the right part of Fig.\ref{likeRat} we show the anti-particle to particle ratios measured at forward rapidity.
The $\pi^{-}/\pi^{+}$ ratio measured at $y=3.1$ is independent of centrality and keeps a value slightly above 1
in agreement with the theoretical models that indicate a small isospin effect at this rapidity.
The $K^{-}/K^{+}$ ratio ($y=2.7$) is also independent of centrality but at a value smaller than at mid-rapidity
presumably due to the high net baryon content which favours the associated production of $K^{+}$.
Contrarily to the dependence shown at mid-rapidity, the $\bar{p}/p$ ratio measured at $y=3.0$ increases
with centrality and is consistent with the larger stopping in central nucleus-nucleus collisions
which shifts the initial protons towards mid-rapidity. The models seem to describe qualitatively well this trend.

\begin{figure}[hpt]
\begin{center}
\includegraphics[height=3.5in]{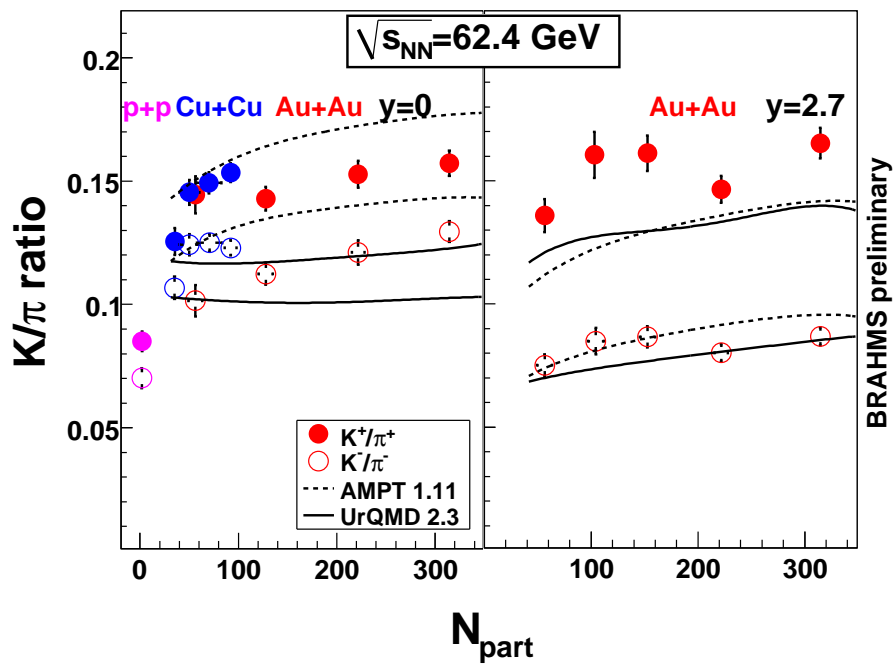}
\end{center}
\caption{$K/\pi$ ratios dependence on the number of participant nucleons at mid-rapidity (left panel)
and forward rapidity (right panel). The data at mid-rapidity includes BRAHMS results from Au+Au (red), Cu+Cu (blue) 
and p+p (magenta) collisions at $\sqrt{s_{NN}}=62.4$ GeV. At forward rapidity only Au+Au results are used.
The curves are model calculations with AMPT (dashed line) and UrQMD (solid line). The upper curve
is for $K^{+}/\pi^{+}$ ratio and the lower curve is for $K^{-}/\pi^{-}$ ratio.}
\label{kpiRat}
\end{figure}

In the left panel of Fig.\ref{kpiRat} we show the $K/\pi$ ratios measured
at mid-rapidity in p+p, Cu+Cu and Au+Au collisions as a function of the number of participant nucleons. 
The error bars are statistical only. 
Both positive and negative measured ratios increase with increasing system size. We also observe that
the $K/\pi$ ratios measured in peripheral Au+Au reactions are smaller than the ones measured
at the same number of participants but in central Cu+Cu collisions. This observation was also made at SPS
energies and interpreted as being due to the different mean number of N+N collisions per projectile 
\cite{spsKpiSystemSize}. 
The theoretical calculations shown are made only for Au+Au collisions
over the entire $\mathrm{N_{part}}$ interval covered by nucleus-nucleus data. 
Thermodynamic models explain the increase of the $K/\pi$ 
ratios with the system size based on the transition from the canonical to grand-canonical
ensemble \cite{statModels}. 
There are also models which suggest that the medium formed after the collision is
made of a thermally equilibrated core and a corona which is a superposition of N+N
sub-collisions in which the strangeness production is suppressed \cite{corecoronaModels}. 
The AMPT and UrQMD theoretical calculations made for Au+Au collisions seem to describe the trend of the data 
but overestimate (AMPT) or underestimate (UrQMD) the absolute values mainly due to disagreements
in the estimation of the pion yields \cite{auau62Yields}.
The forward rapidity $K/\pi$ ratios measured in Au+Au collisions at $y\sim3$ are shown in the 
right panel of Fig.\ref{kpiRat}.
The $K^{+}/\pi^{+}$ and $K^{-}/\pi^{-}$ ratios show the same dependence on centrality as at mid-rapidity
but the negative ratio has much smaller values. The rapidity dependence of these ratios 
has a splitting behaviour towards forward rapidity which was also observed on a smaller scale in Au+Au collisions at
$\sqrt{s_{NN}}=200$ GeV \cite{brahms200Kpi}. 
The theoretical calculations give a good agreement with the $K^{-}/\pi^{-}$ results while 
the $K^{+}/\pi^{+}$ is slightly underestimated.

\section{Conclusions}
The average $p_T$ has been shown to increases with system size at mid-rapidity for all studied particle species and 
being approximately constant at rapidity $y\sim3$. The $\langle p_T \rangle$ calculated for different species 
increases also with the particle rest mass.
The $\pi^-/\pi^+$ and $K^-/K^+$ ratios are independent 
of the number of participants in the covered centrality range both at mid- and forward-rapidity. The $\bar{p}/p$ ratio
at mid-rapidity decreases with increasing system size while at $y\sim3$ it increases which is consistent with
higher baryon stopping in larger colliding systems. Finally, it has been shown that 
the $K/\pi$ ratios are higher in more central collisions both at mid-rapidity and forward rapidity.
The positive and negative $K/\pi$ ratios have a splitting behaviour at forward rapidity
due to the high net-baryon content.
Also, at the same number of participants, the central 
Cu+Cu collisions produce higher amounts of strangeness than peripheral Au+Au collisions.

\end{document}